\documentclass[12pt,letterpaper]{article}
\pdfoutput=1
\usepackage{jcappub}
\usepackage{bm}
\usepackage{amsmath,amssymb}
\usepackage{subfig}
\usepackage{caption}

\newcommand{\Beq}{\begin{equation}\begin{aligned}}
\newcommand{\Eeq}{\end{aligned}\end{equation}}

\usepackage{graphicx, color}
\usepackage{dcolumn}
\usepackage{bm}
\usepackage[mathlines]{lineno}
\usepackage{amsmath}
\usepackage{amssymb}
\usepackage[numbers,sort&compress]{natbib}
\usepackage{bm}
\usepackage{float}
\usepackage{mathrsfs}
\usepackage{lipsum}
\usepackage{hyperref}
\usepackage[normalem]{ulem}
\hypersetup{
    colorlinks=true,       
    linkcolor=rp,          
    citecolor=green2,        
    filecolor=magenta,      
    urlcolor=green2           
}

\usepackage[all]{hypcap} 
\usepackage{cleveref}

\newcommand{\beq}{\begin{align}}
\newcommand{\eeq}{\end\begin{align}}

\def\lap{\lower.5ex\hbox{$\; \buildrel < \over \sim \;$}}
\def\gap{\lower.5ex\hbox{$\; \buildrel > \over \sim \;$}}

\def\be{\begin{align}}
\def\ee{\end\begin{align}}
\def\ba{\begin{eqnarray}}
\def\ea{\end{eqnarray}}

\def\b{\boldsymbol}
\def\bk{\b k}

\def\bp{\b p}

\def\bx{\b x}
\def\tbx{\tilde{\b x}}
\def\by{\b y}

\def\bz{\b z}

\def\bu{\b u}

\def\bv{\b v}

\def\bw{\b w}

\def\sink{\textrm{sinc}}

\newcommand{\fref}[1]{Fig.~\ref{#1}}

\definecolor{rp}{cmyk}{0.2, 1, 0.6, 0}
\definecolor{rp}{cmyk}{0.2, 1, 0.6, 0}
\definecolor{green2}{cmyk}{0.27, 0, 1, 0.52}

\newcommand{\bPsi}{{\bm{\mathsf{\Psi}}}}
\newcommand{\bepsilon}{{\bm{\mathsf{\epsilon}}}}
\newcommand{\bS}{{\bm{S}}}
\newcommand{\bW}{{\bm{W}}}

\title{\huge i-SPin: An integrator for\\
multicomponent Schr\"{o}dinger-Poisson systems with self-interactions}

\author{Mudit Jain,}
\author{Mustafa A. Amin}

\affiliation{Department of Physics and Astronomy, Rice University, Houston, Texas 77005, U.S.A.}

\emailAdd{mudit.jain@rice.edu}
\emailAdd{mustafa.a.amin@rice.edu}

\abstract{We provide an algorithm and a publicly available code to numerically evolve multicomponent Schr\"{o}dinger-Poisson (SP) systems with a SO($n$) symmetry, including attractive or repulsive self-interactions in addition to gravity. Focusing on the case where the SP system represents the non-relativistic limit of a massive vector field, non-gravitational self-interactions (in particular spin-spin interactions) introduce complexities related to mass and spin conservation which are not present in purely gravitational systems. We address them with an analytical solution for the `kick' step in the algorithm, where we are able to decouple the multicomponent system completely. Equipped with this analytical solution, the full field evolution is second order accurate, preserves spin and mass to machine precision, and is reversible. Our algorithm allows for an expanding universe relevant for cosmology, and the inclusion of external potentials relevant for laboratory settings.}

\begin{document}

\maketitle

\tableofcontents

\section{Introduction}
\label{sec:intro}

Physical systems described by Schr\"{o}dinger(-like) equations are ubiquitous in many areas of physics, ranging from (ultra-)light dark matter in cosmology~\cite{Turner:1983he,Press:1989id,Sin:1992bg,Goodman:2000tg,Guzman:2003kt,Amendola:2005ad,Calabrese:2016hmp,Niemeyer:2019aqm,Ferreira:2020fam,Adshead:2021kvl,Jain:2021pnk,Amin:2022pzv,Gorghetto:2022sue,Jain:2022kwq}, to terrestrial systems in non-linear optics~\cite{10.1007/3-540-46629-0_9,menyuk1987nonlinear, christodoulides1988vector, rand2007observation, sun2009bound, baronio2012solutions,1973ZhETF..65..505M,1073308}, water waves~\cite{https://doi.org/10.1002/sapm1967461133,https://doi.org/10.1002/sapm1976553231,doi:10.1146/annurev.fluid.40.111406.102203} and Bose-Einstein condensates (BEC)~\cite{PhysRevLett.81.5257,PhysRevA.79.013423,PhysRevLett.108.010402,Dalfovo:1999zz,2002Natur.417..150S,Nguyen:2014,Luo2020}. Apart from long-range interactions such as gravity in the astrophysical/cosmological context, or external potential in laboratory settings, the Schr\"{o}dinger field can have point-like quartic self-interactions, with its evolution determined by time-dependent Gross–Pitaevskii(-like) equations. 

For a single Schrodinger field, the self-interaction only depends on the number density, and there are many numerical integrators available in the literature. In the astrophysical context, see the algorithms in ~\cite{Springel:2005mi,Schive:2009hw,Mocz:2015sda,Schwabe:2016rze,Mocz:2017wlg,Zhang:2016uiy,Edwards:2018ccc,Nori:2018hud} for simulating the dynamics of ultralight scalar dark matter with only gravitational interaction, and~\cite{Amin:2019ums,Glennon:2020dxs} where quartic self-interaction was also included.

For an $n$-component Schr\"{o}dinger field, however, there are additional self-interaction terms possible, that do not just depend on the number density. When the system respects an SO($n$) symmetry, it is possible to have isospin-isospin interactions, where the total isospin is the conserved charge associated with the SO($n$) symmetry. Such interactions mix the different field components, and the extension of the scalar algorithm towards multicomponent Schr{\"o}dinger field is complex.\footnote{When isospin-isospin interactions are absent, scalar algorithms can be easily extended to evolve multicomponent systems. The time evolution is essentially done for each component separately (still coupled through gravity). For example see~\cite{Amin:2022pzv,Gorghetto:2022sue} for two numerical studies of nonlinear dynamics of vector dark matter with purely gravitational interactions.} We are interested in exploring the nonlinear dynamics associated with such interactions (alongside other interactions), which necessitates developing a numerical algorithm capable of simulating dynamics faithfully. We present a new algorithm that evolves a multicomponent Schr{\"o}dinger-Poisson system with (iso-)spin-spin interaction, that preserves both isospin and mass to machine precision. We also extend the algorithm to allow for dynamics in an expanding universe.

Such multicomponent Schr{\"o}dinger systems with these additional interactions, are present in many areas of physics. For 3-component systems representing vector dark matter, where the aforementioned isospin is the intrinsic spin, spin-spin type interactions  can be naturally present. For instance, they are present in the low-energy effective theory of vector dark matter arising from an Abelian Higgs model (with a heavy Higgs field). These interactions can lead to interesting modifications to the phenomenology of vector dark matter formation \cite{Graham:2015rva,Long:2019lwl,Agrawal:2018vin,Co:2018lka,Dror:2018pdh,Bastero-Gil:2018uel,Co:2021rhi}, black-hole superradiance \cite{Zilhao:2015tya,East:2017ovw,PhysRevD.96.035019,March-Russell:2022zll}, and impact nonlinear small-scale structure in such dark matter \cite{Adshead:2019lbr,Jain:2021pnk, Amin:2022pzv,Gorghetto:2022sue}. As discussed in \cite{Zhang:2021xxa,Jain:2022kwq}, they also remove the energy degeneracy between polarized vector solitons which could potentially impact their cosmological population. Studying such phenomenology, with an eye towards nonlinear spin dynamics, is one of our main motivations.

In a laboratory context, multicomponent/spinor BECs can also include spin-spin type of interactions~\cite{PhysRevA.79.013423,PhysRevLett.108.010402,doi:10.1143/JPSJ.73.2996,PhysRevLett.81.5257}. Our algorithm is equally applicable in this context and can accommodate external potentials such as harmonic traps. There already exists literature on simulating such spin-spin interactions for spin-1 BECs~\cite{PhysRevE.93.053309,CiCP-24-899}, where the three spin multiplicity fields are mixed, and evolved together.\footnote{We thank Han Pu for making us aware of this body of work. See discussion about the key aspects of this approach in section~\ref{sec:nontrivialkick}.} Our algorithm on the other hand, evolves the field in a ``Cartesian" basis where we are able to decouple the different field components. Because of this, we can easily evolve a general Schr{\"o}dinger system with a SO($n$) symmetry.

The organization of this paper is as follows. In section~\ref{Sec:model} we begin by writing down the non-relativistic  Schr\"{o}dinger-Poisson system having both gravitational and point-like self-interactions, lay down the various conserved quantities, and also re-scale the system to work with dimensionless quantities. In section~\ref{sec.integralevolve} we first review the split-step Fourier algorithm, generally employed for evolving a single Schr\"{o}dinger field (scalar system). We then develop an algorithm to evolve vector Schr\"{o}dinger-Poisson system, containing the point-like spin-spin interaction. In section~\ref{sec:numerical.tests} we provide some numerical tests to verify the convergence of our algorithm, its unitary and spin conserving property, along with its time reversibility feature. Then in section~\ref{sec:extension}, we generalize our algorithm to allow for an expanding background as well as evolution of a general $n$-component Schr{\"o}dinger system with a SO($n$) symmetry. Finally in section~\ref{Sec:sum_disc}, we summarize our work. A collection of appendices provide a derivation of the nonrelativistic action, fluid and spin conservation equations, and polarized soliton solutions in vector fields including both gravitational and non-gravitational interactions.

\section{Spin-1 Schr\"{o}dinger-Poisson system}
\label{Sec:model}

We begin with a $3$-component Schr\"{o}dinger-Poisson system with SO($3$) symmetry with non-relativistic massive vector fields in mind. That is, the transformation $\psi_i \rightarrow R_{ij}\psi_j$ (with $R \in$ SO($3$)) of the Schr\"{o}dinger field $\bPsi = (\psi_1,\psi_2,\psi_3)$, leaves the action unchanged.\footnote{We use the Einstein summation convention throughout the paper.} On account of this, we have the following general action that includes both Newtonian gravity and point self-interactions
\begin{align}\label{eq:nonrel_action}
    \!\mathcal{S}_{\rm nr} &\!=\!\! \int \!\!\mathrm dt\,\mathrm d^3x\Biggl[\frac{i\hbar}{2}\bPsi^{\dagger}\cdot\dot{\bPsi} + \mathrm{c.c.} - \frac{\hbar^2}{2m} \nabla\bPsi^{\dagger}\cdot\nabla\bPsi + \frac{1}{8\pi G}\Phi\nabla^2\Phi - m\Phi\bPsi^{\dagger}\cdot\bPsi - V_{\rm nrel}(\bPsi,\bPsi^\dagger)\Biggr].
\end{align}
Here, the first two terms dictate the usual free field evolution (of each of the field component $\psi_i$), while the third and fourth terms account for the Gauss' law for Newtonian gravity where only the mass density $m\bPsi^{\dagger}\cdot\bPsi = m\psi^{\ast}_i\psi_i$ contributes to the Newtonian potential $\Phi$. Finally, the last term accounts for point interactions of the vector field $\bPsi$, and takes the following form for quartic self-interaction
\begin{align}\label{eq:Vnr_vector}
    V_{\rm nrel}(\bPsi^\dagger,\bPsi) &= - \frac{\lambda(\hbar c)^3}{8(mc^2)^2}\Big[(\bPsi\cdot\bPsi)\,(\bPsi^{\dagger}\cdot\bPsi^{\dagger}) + 2\,(\bPsi^{\dagger}\cdot\bPsi)^2\,\Bigr].
\end{align}
In terms of the number density $\rho = \bPsi^\dagger\bPsi$  and spin density $\bm{\mathcal{S}} = i\hbar\bPsi\times\bPsi^{\dagger}$, the spin-spin interaction becomes apparent:
\begin{align}\label{eq:Vnr_vector2}
    V_{\rm nrel}(\rho,\bm{\mathcal{S}}) 
    &= -\frac{\lambda(\hbar c)^3}{8(mc^2)^2}\left[3\rho^2 - \frac{(\bm{\mathcal{S}}\cdot\bm{\mathcal{S}})}{\hbar^2}\right]\,.
\end{align}
This admits the following Schr\"{o}dinger-Poisson system of equations
\begin{align}\label{eq:masterSP1}
    i\hbar\,\partial_t\bPsi &= -\frac{\hbar^2}{2m}\nabla^2\bPsi + m\,\Phi\,\bPsi -  \frac{\lambda(\hbar c)^3}{4(mc^2)^2}\Big[(\bPsi\cdot\bPsi)\,\bPsi^{\dagger} + 2\,(\bPsi^{\dagger}\cdot\bPsi)\,\bPsi\,\Bigr]\,,\nonumber\\
    \nabla^2\Phi &= 4\pi G m\,\bPsi^{\dagger}\cdot\bPsi\,.
\end{align}
The above form of the potential in eq.~\eqref{eq:Vnr_vector} arises from the relativistic quartic potential 
\begin{align}
    V_{\rm rel} = -\lambda(W_{\mu}W^{\mu})^2\,,
\end{align}
upon taking the non-relativistic limit of an effective theory of a self-interacting massive spin-1 field $W_{\mu}$ that is minimally coupled to gravity. See~\cite{Zhang:2021xxa,Jain:2022kwq} for details. For completeness, we also provide salient aspects of this derivation in appendix~\ref{sec:app_nrlimit}.

Such a quartic potential naturally arises in the low energy effective field theory of the Abelian Higgs model, when the heavy Higgs field is integrated out of the spectrum~\cite{Zhang:2021xxa,Jain:2022kwq}.\footnote{Even though such a coupling violates perturbative Unitarity, signaling the need for a proper UV completion (such as the Abelian Higgs model), it can also lead to other problems related to just the classical evolution of the field~\cite{Mou:2022hqb, Clough:2022ygm,Aoki:2022woy}, when
such a proper UV completion is not taken into account. For simulations of Abelian-Higgs model in an expanding universe, which do not encounter any evolution problems in its classical evolution, see \cite{Lozanov:2019jff}.} In this case $\lambda > 0$, dictating attractive self-interaction. This is easily understood on account of the (heavy) scalar particle exchange. While the case of repulsive self-interaction is naturally realized with multiple spin-1 fields (Yang-Mills structure) with appropriate modifications to the form of the potential in~\eqref{eq:Vnr_vector}~\cite{Jain:2022kwq}, in this paper we take a more phenomenological point of view and allow $\lambda$ to also take negative values in~\eqref{eq:Vnr_vector} without including multiple spin-1 fields.

Extension to the FLRW universe can be made by replacing $\nabla\rightarrow \nabla/a$ and $\partial/\partial t\rightarrow \partial/\partial t+3H/2$ in \eqref{eq:masterSP1}. Here $a$ is the scale factor and $H=\dot{a}/a$ is the Hubble parameter. This extension is discussed further in section~\ref{sec:extension}. In the same section, we also discuss more general $n$-component Schr\"{o}dinger-Poisson systems, i.e. $\bPsi = (\psi_1,\psi_2,...,\psi_n)$, with a SO($n$) symmetry. In this case, the factors in front of the two terms in~\eqref{eq:Vnr_vector} can take arbitrary values.

\subsection{Conserved quantities}

The various conserved quantities associated with our non-relativistic system~\eqref{eq:nonrel_action} are
\begin{align}
\label{eq:conserved}
    N&=\int \mathrm d^3x \,\bPsi^\dagger\cdot\bPsi,\quad \textrm{and}\quad M=mN,\qquad\qquad\qquad\quad\,\,\, \textrm{(particle number and rest mass)}\\
    {\b P} &= \frac{\hbar}{2m}\int \mathrm d^3x
    \,\Re\left(i\,\bPsi\cdot\nabla\bPsi^{\dagger}\right)\,,\quad\qquad\qquad\qquad\qquad\qquad\qquad\qquad\quad\, \textrm{(linear momentum)}\\
    E&= \int\mathrm{d}^3x\Biggl[\frac{\hbar^2}{2m}\,\nabla \bPsi^{\dagger}\cdot\nabla \bPsi- \frac{4\pi Gm^2}{2}\bPsi^{\dagger}\cdot\bPsi\int \frac{\mathrm{d}^3y}{4\pi|{\bm{x}}-{\bm{y}}|}\bPsi^{\dagger}({\bm{y}})\cdot\bPsi(\bm{y}) + V_{\rm nrel}\Biggr]\,,\quad \textrm{(energy)}\\
    \bS&= \hbar\int \mathrm d^3x\,i\bPsi\times\bPsi^\dagger\,,\quad\qquad\qquad\qquad\qquad\qquad\qquad\qquad\quad\, \textrm{(spin angular momentum)}\\
    \bm{L}&= \hbar\int \mathrm d^3x
    \,\Re\left(i\,\bPsi^{\dagger}\cdot\nabla\bPsi\times \bx\right). \qquad\qquad\qquad\qquad\qquad\quad\textrm{(orbital angular momentum)}
\end{align}
In general, these conserved quantities find a natural extension for systems with SO$(n)$ symmetry. While for the scalar case ($n=1$) spin is trivially zero, for multicomponent systems ($n > 1$) in general, `spin' is the charge associated with the SO($n$) invariance. See section~\ref{sec:extension} ahead.

In appendix~\ref{sec:app_fluid_conserve_eqns}, we provide local conservation equations for mass and momentum conservation (multicomponent Madelung equations including self-interactions), as well as the continuity equation for spin.

\subsection{Re-scaled system}
\label{sec:rescaled}

In what follows, we shall work with dimensionless quantities. For this purpose, we rescale the fields, space and time in the following fashion
\begin{align}
    t = \frac{\hbar}{\mathcal{E}}\tilde{t} \qquad {\bx} = \frac{\hbar}{\sqrt{m\,\mathcal{E}}}\tbx \qquad \Phi = \frac{\mathcal{E}}{m}\tilde{\Phi} \qquad \psi_i = \left(\frac{\mathcal{E}^2\hbar c}{8\pi Gm}\right)^{1/2}\tilde{\psi}_i\,.
\end{align}
Here, $\hbar/\mathcal{E}$ serves as a measure of the characteristic time-scales present in the system. In terms of these quantities, the Schr\"{o}dinger-Poisson system becomes
\begin{align}\label{eq:rescaled_Schrodinger.Poisson}
    i\partial_{\tilde{t}}\tilde{\psi}_i &= -\frac{1}{2}\tilde{\nabla}^2\tilde{\psi}_i + \tilde{\Phi}\,\tilde{\psi}_i - \tilde{\lambda}\left[\tilde{\psi}_j\tilde{\psi}_j\tilde{\psi}^{\ast}_i + 2\,\tilde{\rho}\,\tilde{\psi}_i\right]\,,\nonumber\\
    \tilde{\nabla}^2\tilde{\Phi} &= \frac{1}{2}\tilde{\rho}\,,
\end{align}
where $\tilde{\rho} \equiv \tilde{\psi}^{\ast}_i\tilde{\psi}_i$ is the re-scaled number density, and
\begin{align}
    \tilde{\lambda} = \frac{\lambda\hbar^4\mathcal{E}}{4m^3\,8\pi G} \approx 0.014 \left(\frac{\lambda}{10^{-84}}\right)\left(\frac{\mathcal{E}/mc^2}{10^{-12}}\right)\left(\frac{10^{-20}\,{\rm eV}}{m}\right)^2\,.
\end{align}
In the description of the algorithm, as well as our numerical code, we use this scaled, dimensionless system of equations~\eqref{eq:rescaled_Schrodinger.Poisson}. The only choices to be made are the value of $\tilde{\lambda}$ and the initial conditions for the field. From now on we shall remove all the tildes to de-clutter our presentation with the understanding that all quantities are dimensionless.

\section{Time evolution}
\label{sec.integralevolve}
We first review the usual algorithm for scalar SP system (including self-interactions), that is employed in the literature. See~\cite{Roulet:2020vqw} for a broad overview of different integrators used for non-linear time-dependent Schr\"{o}dinger type equations. Building upon some of the key concepts discussed from the scalar case, we will develop an algorithm for the vector case that includes point-like self-interactions in subsequent sections.

The presentation below is somewhat formal, and the reader interested in getting to the self-interacting vector algorithm can skip directly to section \ref{sec:vector_algo_summary}.

\subsection{A review of the scalar system}

For the scalar case, the self-interaction  only depends on the number density ($V_{\rm nrel} = -\lambda|\psi^{\ast}\psi|^2$), leading to the following evolution of the Schr\"{o}dinger field
\begin{align}\label{eq:SPsop}
    i\,\partial_t\psi = \mathcal{H}\psi\,\qquad {\rm with} \qquad \mathcal{H} &= \underbrace{-\nabla^2/2} + \underbrace{\Phi -  \lambda\,\rho}\,,\\
    &= \;\;\mathcal{H}_{\rm drift}\;+\;\mathcal{H}_{\rm kick}\nonumber
\end{align}
where  we have defined $\mathcal{H}_{\rm drift}$ and $\mathcal{H}_{\rm kick}$ as `drift' and `kick' Hamiltonians respectively. Explicitly, the Hamiltonian density has a position basis representation $[\mathcal{H}]_{\bx\,\by} 
    = [\mathcal{H}_{\rm drift}]_{\bx\,\by} + [\mathcal{H}_{\rm kick}]_{\bx\,\by}\,,$
where the non-local drift and local (diagonal) kick Hamiltonian densities are
\begin{align}
    [\mathcal{H}_{\rm drift}]_{\bx\,\by} \equiv \frac{1}{2}\int_{\bk}\mathcal{F}^{-1}_{{\bk},{\bx}}\,{\bk}^2\,\mathcal{F}_{{\bk},{\by}}\,,\qquad {\rm and} \qquad [\mathcal{H}_{\rm kick}]_{\bx\,\by} \equiv \left(\Phi(\bx) -  \lambda\,\rho(\bx)\right)\delta^{3}({\bx}-{\by}).
\end{align}
Here, $\mathcal{F}_{{\bk},{\by}} \equiv e^{i{\bk}\cdot{\by}}$ is the Fourier transform matrix element and $\int_{\bk}=\int d^3k/(2\pi)^3$.\footnote{In practice, we work with a finite volume $V=L^3$, for which $\int_{\bk}\rightarrow V^{-1}\sum_{\b k}$ with $\bk=2\pi{\b n}/L$.}  With this representation, the formal solution of \eqref{eq:SPsop} is
\begin{align}\label{eq:psi.correctevolution.scalar}
    \psi(\bx,t + \epsilon) = \int_{\by}U({\bx},{\by},\epsilon)\,\psi({\by},t)\,, \qquad {\rm where} \qquad U({\bx},{\by},\epsilon) &= {\rm T}\,\Bigl[e^{-\frac{i}{\hbar}\int^{t + \epsilon}_{t}\mathrm{d}t'\,\mathcal{H}}\Bigr]_{{\bx}\,{\by}}\,.
\end{align}
Here, ``T" stands for time-ordering and $\int_{\by}=\int d^3y$.\footnote{Note that in our convention there is no $(2\pi)^3$ in the spatial integrals. If space is also discretized then this integral becomes a Reimann sum as usual and $\delta^{(3)}(\bx-\by)=V\delta_{\bx,\by}$.} The evolution operator $U$ satisfies the unitarity relationship
\begin{align}
    \int_{\bz}U^{\dagger}({\bx},{\bz},\epsilon)U({\bz},{\by},\epsilon) = \delta^{3}({\bx}-{\by})\,.
\end{align}
 The unitary evolution operators related to the `drift' and `kick' parts of the Hamiltonian are 
\begin{align}
    &U_{\rm drift}(\bx,\by,\epsilon) \equiv [e^{-i\epsilon\,\mathcal{H}_{\rm drift}}]_{\bx\,\by} = \int_{\bk}\mathcal{F}^{-1}_{\bk,\bx}\;e^{-i\epsilon\,\bk^2/2}\;\mathcal{F}_{\bk,\by}\,,\\
    &U_{\rm kick}(\bx,\by,\epsilon) \equiv [e^{-i\epsilon\,\mathcal{H}_{\rm kick}}]_{\bx\,\by} = e^{-i\epsilon(\Phi(\bx)-\lambda\rho(\bx))}\,\delta^3({\bx}-{\by}).
\end{align}
In terms of $U_{\rm drift}$ and $U_{\rm kick}$, the formal solution in \eqref{eq:psi.correctevolution.scalar}  becomes\footnote{Equivalently, kick-drift-kick set of operations work equally well in so far as $\mathcal{O}(\epsilon^2)$ accuracy, mass conservation, and time reversibility is concerned. Since we'll be using drift-kick-drift set of operations for the vector case ahead, we present the same for the scalar case in order to be consistent in our presentation.}
\begin{align}\label{eq:psi_scalar_full}
    \psi({\bx},t+\epsilon) &= \int_{\bu}\;\;U_{\rm drift}({\bx},{\bu},\epsilon/2)\,\int_{\bv}U_{\rm kick}({\bu},{\bv},\epsilon)\,\int_{\bw}U_{\rm drift}({\bv},{\bw},\epsilon/2)\,\psi({\bw},t) + \mathcal{O}(\epsilon^3)\nonumber\\
    &= \int_{\bp}\mathcal{F}^{-1}_{\bp,\bx}e^{-i\frac{\epsilon}{2}\bp^2/2}\int_{\bu}\mathcal{F}_{\bp,\bu}e^{-i\epsilon(\Phi({\bu})-\lambda\rho({\bu}))}\int_{\bk}\mathcal{F}^{-1}_{\bk,\bu}e^{-i\frac{\epsilon}{2}\bk^2/2}\int_{\bw}\mathcal{F}_{\bk,\bw}\psi({\bw},t) + \mathcal{O}(\epsilon^3).
\end{align}
Note that $\Phi$ and $\rho$ in $\mathcal{H}_{\rm kick}$ are evaluated after the first half-drift is performed. This is crucial for accuracy of $\mathcal{O}(\epsilon^2)$ and to maintain time-reversal symmetry. 

To see this reversibility in effect, let us first evolve the system forward for half a time step of $\epsilon/2$ by using $U_{\rm drift}$ and $U_{\rm kick}$:
\begin{align}\label{eq:psi_scalar_forw_1/2}
    \psi({\bz},t + \epsilon/2) &= \int_{\bv}\;\;U_{\rm kick}({\bz},{\bv},\epsilon/2)\,\int_{\by}U_{\rm drift}({\bv},{\by},\epsilon/2)\,\psi({\by},t) + \mathcal{O}(\epsilon^2)\nonumber\\
    &= e^{-i\frac{\epsilon}{2}(\Phi(\bz)-\lambda\rho(\bz))}\,\int_{\bk}\mathcal{F}^{-1}_{\bk,\bz}\;e^{-i\frac{\epsilon}{2}\,\bk^2/2}\;
    \int_{\by}\mathcal{F}_{\bk,\by}\,\psi({\by},t) + \mathcal{O}(\epsilon^2)\,,
\end{align}
where $U_{\rm kick}$ is evaluated using the field obtained after the half-drift has been performed. In conjugation to this, let us now evolve the system backwards from $t + \epsilon$ to $t + \epsilon/2$ using time reversed operations:
\begin{align}\label{eq:psi_scalar_back_1/2}
    \psi({\bz},t + \epsilon/2) &= \int_{\bv}\,U'_{\rm kick}({\bz},{\bv},-\epsilon/2)\,\int_{\bx}U_{\rm drift}({\bv},{\bx},-\epsilon/2)\,\psi({\bx},t + \epsilon) + \mathcal{O}(\epsilon^2)\nonumber\\
    &= e^{i\frac{\epsilon}{2}(\Phi'(\bz)-\lambda\rho'(\bz))}\,\int_{\bk}\mathcal{F}^{-1}_{\bk,\bz}\;e^{i\frac{\epsilon}{2}\,\bk^2/2}\;\int_{\bx}\mathcal{F}_{\bk,\bx}\,\psi({\bx},t+\epsilon) + \mathcal{O}(\epsilon^2)\,.
\end{align}
Here too, $U'_{\rm kick}$ is evaluated using the field after the half-drift has been performed. We can invert the above eq.~\eqref{eq:psi_scalar_back_1/2} to get
\begin{align}\label{eq:psi_scalar_back_inverted_1/2}
    \psi({\bx},t + \epsilon) &= \int_{\bv}U_{\rm drift}^{-1}({\bv},{\bx},-\epsilon/2)\,\int_{\bz}U^{'\,-1}_{\rm kick}({\bz},{\bv},-\epsilon/2)\,\psi({\bz},t + \epsilon/2) + \mathcal{O}(\epsilon^2)\nonumber\\
    &= \int_{\bk}\mathcal{F}^{-1}_{\bk,\bx}\;e^{-i\frac{\epsilon}{2}\,\bk^2/2}\;\int_{\bv}\mathcal{F}_{\bk,\bv}\,e^{-i\frac{\epsilon}{2}(\Phi'(\bv)-\lambda\rho'(\bv))}\,\psi({\bv},t+\epsilon/2) + \mathcal{O}(\epsilon^2)\,,
\end{align}
and use it in~\eqref{eq:psi_scalar_forw_1/2} to get the full unitary evolution from $t$ to $t+\epsilon$:
\begin{align}\label{eq:psi_scalar_full2}
    \psi({\bx},t+\epsilon) &=\int_{\bu}\!U_{\rm drift}^{-1}({\bx},{\bu},-\epsilon/2)\int_{\bv}\!U^{'\,-1}_{\rm kick}({\bu},{\bv},-\epsilon/2)\int_{\bw}
    \!U_{\rm kick}({\bv},{\bw},\epsilon/2)\int_{\by}\!U_{\rm drift}({\bw},{\by},\epsilon/2)\psi({\by},t)
\nonumber\\
    &\qquad\qquad\qquad\qquad\qquad\qquad\qquad\qquad\qquad \qquad\qquad\qquad\qquad+ \mathcal{O}(\epsilon^3)\,.
\end{align}
Now note that since each of the exponential operators is time reversible, i.e. $U_{\rm kick/drift}^{-1}(-\epsilon) = U_{\rm kick/drift}(\epsilon)$, the above expression reduces to eq.~\eqref{eq:psi_scalar_full}, proving that the algorithm is indeed time reversible. Furthermore, the evolution accuracy is now $\mathcal{O}(\epsilon^2)$ and the error is $\mathcal{O}(\epsilon^3)$. This can be seen by noting that the re-evaluation of the kick Hamiltonian after the drift step takes into account the leading order correction due to the time integral $\int^{t+\epsilon}_{t}\mathrm{d}t'\, \mathcal{H}(t')$ in ~\eqref{eq:psi.correctevolution.scalar}. See~\cite{Edwards:2018ccc} for some details.

It is worth noting that for the case under consideration, the kick operator only rotates the phase of the field at every point in space, while itself only depending upon the norm $\rho$. As a result, the kick operator doesn't change with time during the kick operation. That is, it is manifestly true that $U_{\rm kick}^{'\,-1}(\bu,\bv,-\epsilon/2)$ evaluated using the backward half-drifted field from $t+\epsilon$, is equal to $U_{\rm kick}(\bu,\bv,\epsilon/2)$ evaluated using the forward half drifted field from $t$.

\subsubsection{Algorithm summary}
Starting with the field $\psi({\bx},t)$ at time $t$, it is evolved through a time step $\epsilon/2$ according to the differential equation
\begin{align}\label{eq:scalar_step1}
    i\partial_{t}\psi = -\frac{1}{2}\nabla^2\psi \qquad {\rm giving} \qquad \psi^{(1)}({\bx}) = \int_{
    \bk}\mathcal{F}^{-1}_{\bk,\bx}\;e^{-i\,\epsilon\,\bk^2/4}\int_{\bw}\mathcal{F}_{\bk,\bw}\,\psi({\bw},t)\,.
\end{align}
Then, it is evolved through a time step $\epsilon$ according to the differential equation
\begin{align}\label{eq:scalar_step2}
    i\partial_{t}\psi = (\Phi - \lambda\rho)\psi \qquad {\rm giving} \qquad \psi^{(2)}({\bx}) = e^{-i\,\epsilon(\Phi^{(1)}(\bx) - \lambda\rho^{(1)}(\bx))}\psi^{(1)}({\bx})\,,
\end{align}
where $\Phi^{(1)}(\bx) = (1/2)\nabla^{-2}_{\bx}(\rho^{(1)}(\bx) - \bar{\rho}^{(1)})$ and overbar represents spatial average.\footnote{In practice, the Newtonian potential is computed using Fourier transformation: $\Phi(\bx) = -\int_{\bk}\mathcal{F}^{-1}_{\bk,\bx}\;\frac{\rho(\bk)}{2\bk^2}$ (without the $\bk = 0$ mode).} Finally, it is again evolved using the following differential equation through a time step $\epsilon/2$
\begin{align}\label{eq:scalar_step3}
    i\partial_{t}\psi = -\frac{1}{2}\nabla^2\psi \qquad {\rm giving} \qquad \psi({\bx},t+\epsilon) = \int_{\bk}\mathcal{F}^{-1}_{\bk,\bx}\;e^{-i\,\epsilon\,\bk^2/4}\int_{\bw}\mathcal{F}_{\bk,\bw}\,\psi^{(2)}({\bw})\,.
\end{align}
Note that for every differential evolution above, the corresponding integral evolution is exact. While the half drift steps result in $\mathcal{O}(\epsilon^2)$ accuracy in the set of drift-kick-drift operations, the successive computation of the kick (after the performance of drift) ensures reversibility. Every operation being unitary ensures unitary evolution throughout.

\subsection{Vector system}

For the vector case, the differential evolution takes a matrix form 
\begin{align}\label{eq:H_noSI_ini}
    i\,\partial_t\psi_i = [\mathcal{H}]_{ij}\psi_j \qquad {\rm with} \qquad [\mathcal{H}]_{ij} &= \underbrace{-\delta_{ij}\,\nabla^2/2} + \underbrace{\delta_{ij}(\Phi -  2\lambda\,\rho) - \lambda\,\psi^{\ast}_i\psi_j}\,.\\
    &= \quad [\mathcal{H}_{\rm drift}]_{ij} + \qquad\quad [\mathcal{H}_{\rm kick}]_{ij}\nonumber
\end{align}
In particular, note the non-diagonal piece, $\propto \lambda\psi^{\ast}_i\psi_j$, in $[\mathcal{H}_{\rm kick}]_{ij}$. The Hamiltonian density, written explicitly in the position basis, is
\begin{align}
    [\mathcal{H}]_{\bx\,\by\,;\,ij} &= \left[\frac{1}{2}\int_{\bk}\mathcal{F}^{-1}_{{\bk},{\bx}}\,{\bk}^2\,\mathcal{F}_{{\bk},{\by}} + \left(\Phi(\bx) -  2\lambda\,\rho(\bx)\right)\delta^{3}({\bx}-{\by})\right]\delta_{ij} - \lambda\,\psi^{\ast}_i(\bx)\psi_j(\bx)\,\delta^{3}({\bx}-{\by})\,,\nonumber
\end{align}
and as always, we have suppressed the time dependence of the fields. Here the latin indices run over the field components, with the (unitary) integral evolution for this system being
\begin{align}\label{eq:psi.correctevolution.vector}
    \psi_i({\bx},t + \epsilon) = \int_{\by}\,U_{ij}({\bx},{\by},\epsilon)\,\psi_j({\by},t) \quad {\rm with} \quad U_{ij}({\bx},{\by},\epsilon) &= {\rm T}\,\Bigl[e^{-\frac{i}{\hbar}\int^{t+\epsilon}_{t}\mathrm{d}t'\,\mathcal{H}(t')}\Bigr]_{\bx\,\by\,;\,ij}\,.
\end{align}
The unitary condition is the following
\begin{align}
    \int_{\bz}\;U_{ik}^{\dagger}({\bx},{\bz},\epsilon)\,U_{kj}({\bz},{\by},\epsilon) = \delta_{ij}\,\delta^{3}({\bx}-{\by})\,.
\end{align}
With these definitions at hand, we shall employ the previously discussed algorithm for scalars, with appropriate modifications appearing due to the non-diagonal kick in the vector SP system.

\subsubsection{Algorithm summary}
\label{sec:vector_algo_summary}

The algorithm proceeds as follows. Starting with the field components $\psi_i({\bx},t)$ at time $t$, they are `drifted' through a time step $\epsilon/2$ according to
\begin{align}\label{eq:vector_step1}
    i\partial_{t}\psi_i = -\frac{1}{2}\nabla^2\psi_i \qquad {\rm giving} \qquad \psi^{(1)}_i({\bx}) = \int_{\bk}\mathcal{F}^{-1}_{\bk,\bx}\;e^{-i\epsilon\,\bk^2/4}\;\int_{\bw}\mathcal{F}_{\bk,\bw}\psi_i({\bw},t)\,.
\end{align}
Then, every component is `kicked' through a time step $\epsilon$ according to
\begin{align}\label{eq:vector_step2}
    i\partial_{t}\psi_i = [\mathcal{H}_{\rm kick}]_{ij}\psi_j \qquad {\rm giving} \qquad \psi^{(2)}_{i}({\bx}) = [U_{\rm kick}(\epsilon)]_{ij}\,\psi^{(1)}_j({\bx})\,,
\end{align}
where recall that $[\mathcal{H}_{\rm kick}]_{ij} = \delta_{ij}(\Phi -  2\lambda\,\rho) - \lambda\,\psi^{\ast}_i\psi_j$, and $[U_{\rm kick}(\epsilon/2)]_{ij}={\rm T}\exp[-i\int_{t}^{t+\epsilon}\mathcal{H}_{\rm kick}(t')]_{ij}$. The operator $U_{\rm kick}$ is evaluated using the field $\psi^{(1)}_{i}$; its evaluation based on an analytic solution of the kick equation is the key step, which we discuss in detail in the next subsection. The explicit expression for $U_{\rm kick}$ is provided in~\eqref{eq:kick_operator}. Once this operation is performed, the fields are again drifted through a time step $\epsilon/2$
\begin{align}\label{eq:vector_step3}
    i\partial_{t}\psi_i = -\frac{1}{2}\nabla^2\psi_i \qquad {\rm giving} \qquad \psi_i({\bx},t+\epsilon) = \int_{\bk}\mathcal{F}^{-1}_{\bk,\bx}\;e^{-i\epsilon\,\bk^2/4}\int_{\bw}\mathcal{F}_{\bk,\bw}\,\psi^{(2)}_j({\bw})\,.
\end{align}
The half drift steps in the set of operations ensure $\mathcal{O}(\epsilon^2)$ accuracy, while successive computation of the kick ensures reversibility. Since every operation is unitary, the algorithm conserves total mass. It also conserves total spin. 

\subsubsection{Kick step}
\label{sec:nontrivialkick}
We want an exact solution for the kick operation $[U_{\rm kick}]_{ij}\psi_j$, in order to ensure reversibility along with mass and spin conservation. This was trivial for the scalar and vector case with no spin-spin interactions since $[\mathcal{H}_{\rm kick}]_{ij}$ was a constant of motion throughout the kick step. In that case the solution for the kick step was the exponential of the kick Hamiltonian density. With self-interactions, due to the non-diagonal piece in $\mathcal{H}_{\rm kick}$, the evolution of each of the field component $\psi_i$ becomes convoluted via the mixing of the field components and non-linearity of the kick differential equation. In this section we work with the kick step governed by the following differential equation
\begin{align}\label{eq:vector-kick}
    i\partial_t\psi_i= [\delta_{ij}(\Phi-2\lambda\rho) -\lambda \psi_i^*\psi_j]\psi_j\,.
\end{align}
Working with the vector system, we first quickly outline an approach where the solution to the above 3-level system is obtained by exponentiating a time-independent $3 \times 3$ matrix. Upon adding and subtracting $\lambda\rho\psi_i = (\lambda\psi_i\psi^{\ast}_j)\psi_j$ in the right hand side, we note that the off-diagonal term $\propto \psi_i\psi^{\ast}_j - \psi_i^*\psi_j$ is the skew-symmetric Hermitian matrix $M_{ij} = -i\varepsilon_{ijk}\mathcal{S}_k$, where $\mathcal{S}_j = i\,\varepsilon_{jkl}\psi_k\psi^{\ast}_l$ is the spin density.\footnote{The relevant equation of motion can also be obtained by varying the potential written in the form~\eqref{eq:Vnr_vector2} directly.} It can be easily seen that both the spin density and the number density (hence also the gravitational potential $\Phi$) are conserved throughout the kick step.\footnote{This is also manifest from the continuity equations in appendix~\ref{sec:app_fluid_conserve_eqns}, since in this step there are no flow of currents.} Hence, the matrix $M$ is constant, rendering the following simple solution
\begin{align}
    \psi_i(t) = [U_{\rm kick}(t-t_0)]_{ij}\psi_j(t_0) \quad {\rm with} \quad [U_{\rm kick}(\epsilon)]_{ij} = e^{-i(\Phi - 3\lambda\rho)\,\epsilon}\,[e^{-i\lambda M \,\epsilon}]_{ij}\,,
\end{align}
where matrix exponentiation is understood, and has the following specific entries
\begin{align}
    [e^{-i \lambda M \epsilon}]_{ij} = \frac{1}{\mathcal{S}^2}\begin{cases}
        \mathcal{S}_i^2 + (\mathcal{S}^2-\mathcal{S}_i^2)\cos\lambda \mathcal{S}\epsilon & i = j \\
        \mathcal{S}_i\mathcal{S}_j(1-\cos\lambda \mathcal{S}\epsilon) - \epsilon_{ij k}\mathcal{S}_k\mathcal{S}\sin\lambda \mathcal{S}\epsilon & i, j, k\quad \textrm{distinct}
    \end{cases}
\end{align}
where no summation is assumed, and $\mathcal{S}$ is the magnitude of the spin density. Another complimentary way of evolving the system is to work in the spin basis where the field is decomposed into the three different spin multiplicities~\cite{Jain:2021pnk} (as opposed to working with Cartesian components as we did here). For example with $\bPsi = \sum_{\lambda = -1,0,1}\,\psi_{\lambda}\,\bepsilon^{(\lambda)}_{\hat{z}}$ where $\bepsilon^{(0)}_{\hat{z}} = (0,0,1)^{T}$ and $\bepsilon^{(\pm 1)}_{\hat{z}} = (1, \pm i, 0)^{T}/\sqrt{2}$, the evolution of ${\b \chi}' = (\psi_{1},\psi_{0},\psi_{-1})$ due to the off-diagonal kick step becomes $i\partial_t{\b \chi}' = \lambda P\,{\b \chi}'$, where $P$ is also a $3 \times 3$ constant skew-symmetric Hermitian matrix. This gives the desired solution ${\b \chi}'(t) = \exp[-i\lambda P(t-t_0)]{\b \chi}'(t_0)$. See~\cite{PhysRevE.93.053309} for this approach.\\

In the scheme above, the different field components are mixed due to (iso-)spin-spin coupling. As a result, the expression for $M_{ij}$ for an $n$-component system can become convoluted as $n$ increases (see seection~\ref{sec:extension}). Below, we continue with the vector system and present a new solution for the above kick differential equation, in which all the Cartesian field components are decoupled. Only the real and imaginary parts of each Cartesian component remain coupled. This decoupling yields a much simpler evolution for each component, and allows for a straightforward generalization to $n$-component Schr{\"o}dinger systems (given in section~\ref{sec:extension}).

We begin with the ansatz $\psi_j(t) = e^{-i(\Phi - 2\lambda\rho)(t-t_0)}\chi_j(t)$ in~\eqref{eq:vector-kick} to get non-linear differential equations for $\chi_j$ with only the non-diagonal piece in $\mathcal{H}_{\rm kick}$ contributing to its evolution:
\begin{align}\label{eq:ODE_nd}
    i\partial_t\chi_i = -\lambda\,\chi^{\ast}_i\,\chi_j\chi_j\,. 
\end{align}
Then, upon multiplying the above with $\chi_i$ and vector summing, we get $i\partial_t(\chi_j\chi_j) = -2\lambda\rho \chi_j\chi_j$. This equation has a simple solution
\begin{align}
    \chi_j\chi_j(t) = re^{i\theta+2i\lambda\rho (t-t_0)}
\end{align}
where $r$ is the magnitude of $\chi_j\chi_j$ (a constant of motion)\footnote{This can also be seen by noting that $r^2 = \rho^2-\mathcal{S}_i\mathcal{S}_i$, and since both $\rho$ and $\mathcal{S}_i$ are constants of motion throughout the kick evolution, so is $r$.}, and $\theta$ is its initial phase (at $t=t_0$). Owing to this analytical (and simple) evolution of the squared sum of the field components, we can now decouple the different field components by using it in~\eqref{eq:ODE_nd}. This renders the following linear differential equation for each of the field components
\begin{align}\label{eq:ndkick_decoupled}
    i\partial_t\chi_i = -\lambda\,r\,e^{i\theta+2i\lambda\rho (t-t_0)}\,\chi^{\ast}_i.
\end{align}
In terms of the real and imaginary parts of $\chi_i$ : $u^i_1 = \Re[\chi_i]$ and $u^i_2 = \Im[\chi_i]$, this differential equation becomes 
\begin{align}\label{eq:ODE.uandv}
    \frac{\partial}{\partial\eta}
    u = (A\,{\b b}\cdot{\vec{\sigma}})
    \,u \quad {\rm with} \quad A = \frac{r}{2\rho}\,,\quad
     {\b b} = (\cos\eta,\, 0,\, -\sin\eta)\,,\quad u=(u_1,u_2)\,.
\end{align}
We have dropped the vector component index on $u$ (related to $``i"$ in $\chi_i$), since each $\chi_i$ now evolves the same way,  differing only in their initial condition.
We have also re-scaled the time as $\eta \equiv \theta + 2\lambda\rho (t-t_0)$ for convenience (with $\eta_0 = \theta$), and $\sigma_i$ are the (real) Pauli matrices.\footnote{Our convention is $\sigma_{1} =
    \begin{pmatrix}
     0 && 1\\
     1 && 0\,
    \end{pmatrix}\;;\;
    \sigma_{2} =
    \begin{pmatrix}
     0 && 1\\
     -1 && 0\,
    \end{pmatrix}\;;\;
    \sigma_{3} =
    \begin{pmatrix}
     1 && 0\\
     0 && -1\,
    \end{pmatrix}$.} 
This is the equation for which we seek an analytical solution.\footnote{Interestingly, this equation can be mapped to the problem of an electrically charged spin-$1/2$ fermion (at rest) in a rotating magnetic field. This can be achieved by making $r/2\rho$ imaginary, with its magnitude representing the magnetic moment of the fermion, and identifying $u$ as the two component Dirac spinor.}\\

We note that the time dependence of the matrix ${\b b}\cdot\vec{\sigma}$ can be stripped off by going to a rotating frame. The corresponding transformation is achieved via the matrix $P = \exp[-\eta\,\sigma_2/2]$. Then, with $v = P^{-1}u$ and $P^{-1}({\b b}\cdot\vec{\sigma})P = \sigma_1$, we get the following transformed equation with the corresponding trivial solution
\begin{align}\label{eq:ODE.v1andv2}
    \frac{\partial}{\partial\eta}
    v 
    &= (A\,\sigma_1 + \sigma_2/2)
    v\, \implies v(\eta) = \exp\Bigl[(\eta-\eta_0)\,(A\,\sigma_1 + \sigma_2/2)\Bigr]v(\eta_0)\,.
\end{align}
Using $u=P v$ and re-instating quantities in terms of the `actual' time $t$, we therefore have the following solution for $u$
\begin{align}\label{eq:u_solution}
    u(t) 
    & = \exp\Bigl[- \lambda\rho(t-t_0)\,\sigma_2\Bigr]\,\exp\Bigl[-\theta\sigma_2/2\Bigr]\,\exp\Bigl[2\lambda\rho(t-t_0)\,(A\,\sigma_1 + \sigma_2/2)\Bigr]\exp\Bigl[\theta\sigma_2/2\Bigr]u(t_0)\,,\nonumber\\
    & \equiv \mathcal{U}(t-t_0)u(t_0)\,.
\end{align}
With this exact non-diagonal kick evolution, the real and imaginary parts of each of the field component $\chi_i$ are evolved using $\mathcal{U}(t-t_0)$, accompanied by a phase translation $e^{-i(\Phi-2\lambda\rho)(t-t_0)}$ due to the diagonal Hamiltonian density. In terms of the field $\psi_j$ itself, the evolution is
\begin{align}\label{eq:kick_operator}
    [U_{\rm kick}(\epsilon)]_{ij}\,\psi_j = \delta_{ij}e^{-i(\Phi - 2\lambda\rho)\epsilon}\Biggl[\Bigl[\mathcal{U}(\epsilon)_{11} + i\,\mathcal{U}(\epsilon)_{21}\Bigr]\Re[\psi_j] + \Bigl[\mathcal{U}(\epsilon)_{12} + i\,\mathcal{U}(\epsilon)_{22}\Bigr]\Im[\psi_j]\Biggr]\,,
\end{align}
where we used $t-t_0=\epsilon$.
Including the drift evolution, the full vector field is therefore evolved according to steps~\eqref{eq:vector_step1},~\eqref{eq:vector_step2}, and~\eqref{eq:vector_step3}, with the above solution for $U_{\rm kick}$. In Appendix~\ref{sec:app:Uexplicit}, we provide the explicit expression for $\mathcal{U}$. This fetches our desired mass and spin conserving, time-reversible drift + kick + drift algorithm, to evolve the multicomponent/vector SP system. Since both the drift and kick evolutions are now exact, the overall accuracy of the system as compared to the true evolution in~\eqref{eq:psi.correctevolution.vector}, like the scalar case, is $\mathcal{O}(\epsilon^2)$.\\

Let us briefly comment on the operator $\mathcal{U}$ in regards to the conservation of mass and spin density
\begin{align}\label{eq:rho_spin_U}
    \rho = u^i_c\,\mathcal{U}^{T}_{ca}\,\,\mathcal{U}_{ab}\,u^i_b \qquad {\rm and} \qquad
    \mathcal{S}_i = \varepsilon_{ijk}(u^j_1\,u^k_2 - u^k_1\,u^j_2)\det\,\mathcal{U}\,.
\end{align}
While $\det\,\mathcal{U} = 1$ indicates that the spin density remains unchanged throughout the kick evolution, $\mathcal{U}^{T} \neq \mathcal{U}^{-1}$ indicates that the mass density within \textit{each} component is not conserved. This is reflective of the fact that there is no separate U($1$) symmetry within each component of the vector field, and hence no mass conservation within each component. On the other hand, the total mass density is conserved. This may seem peculiar since $\mathcal{U}^{T} \neq \mathcal{U}^{-1}$. We note that since $r$, $\rho$ and $\theta$ appearing in the operator are consistent with the initial $\psi_i$ (meaning $\psi_j\psi_j(t_0) = r\,e^{i\theta}$ and $\psi^{\ast}_j\psi_j(t_0) = \rho$), the total mass density $\rho$ is indeed conserved. In general, when this consistency doesn't hold, $\mathcal{U}$ is not $\rho$ conserving.

\subsubsection{Courant–Friedrichs–Lewy condition}\label{sec:cfl.condition}

The Courant–Friedrichs–Lewy (CFL) condition ensures that the time step $\epsilon$ is sufficiently small in order to resolve the fastest process happening in the simulation. For the drift-kick-drift evolution of the multi-component Schr\"{o}dinger system, this means resolving the fastest kick and drift processes every time step. The drift evolution is governed by the operator $U_{\rm drift} \sim e^{i(\epsilon/2)\nabla^2/2}$ dictating that $\epsilon$ be at-least as small as $2\pi(\Delta x)^{2}/3$ in order to resolve one full rotation of the drift phase.\footnote{Here, $\nabla^2$ gets replaced by $\sum^{3}_{i=1}(\Delta x)^{-2}\times 4\sin^2(n_i\pi/N)$ on a discrete lattice, and we set $n=N/2$ along with $\sum \rightarrow 3$ in order to maximize the sum over sine functions.} For the diagonal kick step, the operation is governed by $\sim e^{-i\epsilon (\Phi-2\lambda\,\rho)} \sim e^{-i\epsilon\Phi}\,e^{2i\epsilon\lambda\,\rho}$ which requires that $\epsilon$ be at least as small as the smallest of the two quantities $2\pi|\Phi|^{-1}$ and $2\pi|2\lambda\,\rho|^{-1}$. Finally, the non-diagonal kick dictates that epsilon be at least as small as $2\pi|\lambda|^{-1}\times \min [\rho^{-1}, r^{-1}]$ in order to resolve both the $\sigma_1$ and $\sigma_2$ containing parts of the exponentials appearing in $\mathcal{U}$ in the previous section. Since $r < \rho$ by definition, this condition is less constraining than the one appearing due to diagonal kick. With this, all of the above requirements together result in the following CFL condition:\footnote{This is the same as that for a single component (scalar) case, with appropriate rescaling of $\lambda$.}
\begin{align}\label{eq:CFL.cond}
    \epsilon &= 2\pi\, \delta\,\min\Biggl[\frac{1}{3}(\Delta x)^{2} \,,\, |\Phi|^{-1} \,,\, |2\lambda\,\rho|^{-1}]\Biggr] \quad {\rm with} \quad \delta \ll 1\,.
\end{align}
In our simulations we take $\delta \sim 1/15$, i.e. we resolve the fastest oscillation in the system by $\sim 15$ points within its full $2\pi$ cycle.

With the above CFL condition, the time step $\epsilon$ can be driven to small values when sufficiently dense regions (with $|\lambda|\rho \gtrsim \{|\Phi|,(\Delta x)^{-2}\}$) start to appear in the course of evolution of the Schr{\"o}dinger field. Care should be taken in interpreting results in this regime since, in general, large variations in $\rho$ over sufficiently small spatial regions can take us out of the domain of our non-relativistic theory. Typically in the repulsive case, $\rho$ does not develop large enough variations to endanger our nonrelativistic approximations. However, the situation in more precarious in the attractive self-interaction case.

Consider a spatial region of size $r$ and total enclosed mass $M$, with a roughly isotropic mass density around it.  The total energy in the region $E_{\rm tot}$ is due to contributions from gradient pressure $E_{\nabla} \sim \hbar^2M/(mr)^2$, gravity $E_{\rm grav} \sim -GM^2/r$, and self-interaction $E_{\rm self} \sim -\lambda\,\hbar^3 M^2/(m^4c\,r^3)$. It can be seen that as $|E_{\rm self}+E_{\rm grav}| > E_{\nabla}$ (meaning $E_{\rm tot} < 0$), there is a runaway possibility (when the attractive self-interaction is relevant) where the system can keep on lowering its energy by either focusing all this mass to ever-smaller regions, and/or by accumulating more mass from its surroundings in a given region. 
This process can lead to very large densities in small regions of space.\footnote{In practice, density fluctuations can only increase to a certain extent due to the absence of wavenumbers larger than $\sim \Delta x$ on a discrete lattice. Nevertheless, formation of such `crunched' regions is indicative of this runaway scenario, and hence a breakdown of the nonrelativistic EFT.}
At this point, relativistic corrections (including higher-order terms in the self-interaction which are not present in the system being simulated) cannot be justifiably  ignored. Note that this discussion equally applies to a single component SP system (scalar) with point self-interactions, and is not particular to multicomponent systems only. For the demonstration of fidelity of our algorithm/code, we pick $\Delta x$ small enough so that ${\rm min}[\hdots]=(\Delta x)^2/3$ in the CFL condition throughout the duration of the simulation. 

\section{Numerical tests and results}\label{sec:numerical.tests}

To test our algorithm, we ran several simulations including: (1) a single sitting soliton with different polarizations, (2) two, three or more soliton collisions,  for $\lambda > 0$ and $\lambda < 0$. For a given value of $\lambda$, the spherically symmetric soliton configurations were obtained by numerical shooting in~\eqref{eq:TISP}, which were then put on the discrete lattice. See appendix~\ref{sec:solitons} for a brief discussion of vector solitons arising in a self-interacting massive spin-1 field.

For illustrating the robustness of our algorithm we present results from a set of both forward and time reversed test simulations of a three soliton collision scenario. In these simulations, we worked with a $81^3$ (periodic) grid of dimensionless length $L = 25$, with $\lambda = \pm 0.01$. All three solitons had a $95\%$ radius $R_{\rm sol} \approx 3.7$, with two of them linearly polarized, and one circularly polarized.

We reserve the investigation of more involved scenarios such as many soliton collisions, emergence of solitons from random initial conditions etc., for a separate work.

\subsection{Mass and Spin conservation}\label{sec:reverse}

\begin{figure}[t!]
    \centering
      \includegraphics[width=6.0in]{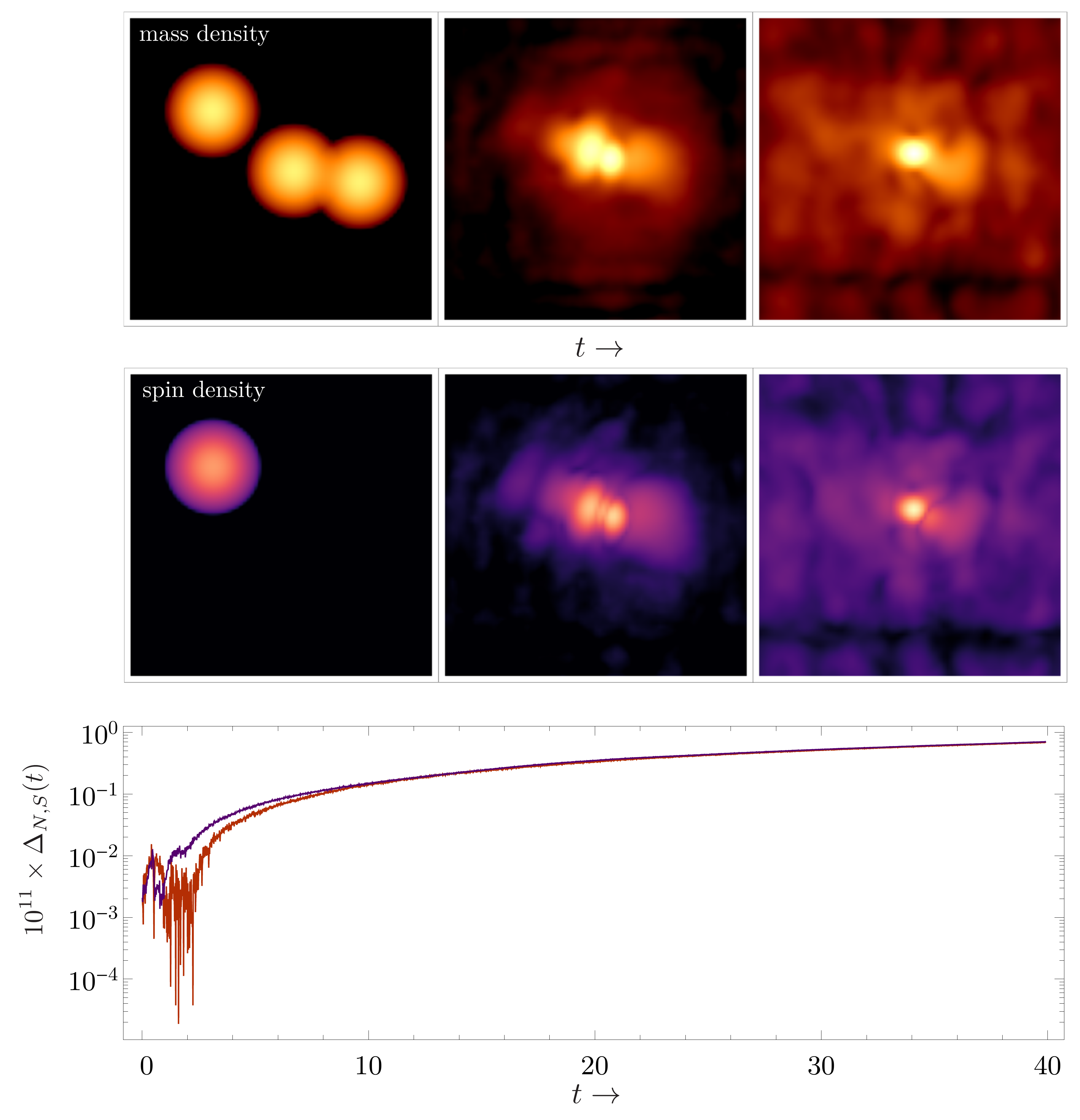}
\caption{Mass and spin conservation: Top panel shows snapshots of projected mass density at three instants $t = 0, 13$, and $40$ (upper panel).  Lower panel are snapshots of magnitude of spin density at the same times. One of the solitons is initialized with maximal spin, whereas the other two have zero spin initially. The bottom plot shows quantitative measures of total spin (blue curve) and total mass (red curve) conservation; both are conserved to better than one part in $10^{11}$. Animations of the simulation results can be seen on~\href{https://www.youtube.com/watch?v=KUbVtULiBtE}{Youtube}.}
      \label{fig:MassSpin}
\end{figure}

Since every step in the drift-kick-drift operation is both mass and spin conserving, the overall evolution is guaranteed to be unitary and spin conserving. We track the fractional change in the total mass and spin:
\begin{align}
    \Delta_{N}(t) = \frac{|N(t)-N(0)|}{N(0)}\,,\qquad\qquad\,
    \Delta_{S}(t) = \frac{1}{3}\sum^{3}_{i=1}\frac{|S_i(t)-S_i(0)|}{|S_i(0)|}\,,
\end{align}
which is expected to be zero up to machine precision. Note that the fractional change in the spin along any direction only makes sense if it was nonzero to begin with.\footnote{There are other possible measures to track the conservation of spin. For example if at-least one of the spin components is not zero, one can calculate $\left(|{\b S}(t) - {\b S}(0)|\right)\left(|{\b S}(0)|\right)^{-1}$.} The plot at the bottom of~\fref{fig:MassSpin} show these two quantities for our chosen three soliton collision scenario (for $\lambda = 0.01$). Note that mass and spin are conserved to machine precision.\footnote{The apparent linear growth of machine level errors in total mass and spin, is likely due to the implementation of fast Fourier transform. We have observed this peculiarity in both Python and Mathematica.}

\begin{figure}[t!]
    \centering
      \includegraphics[width=6.0in]{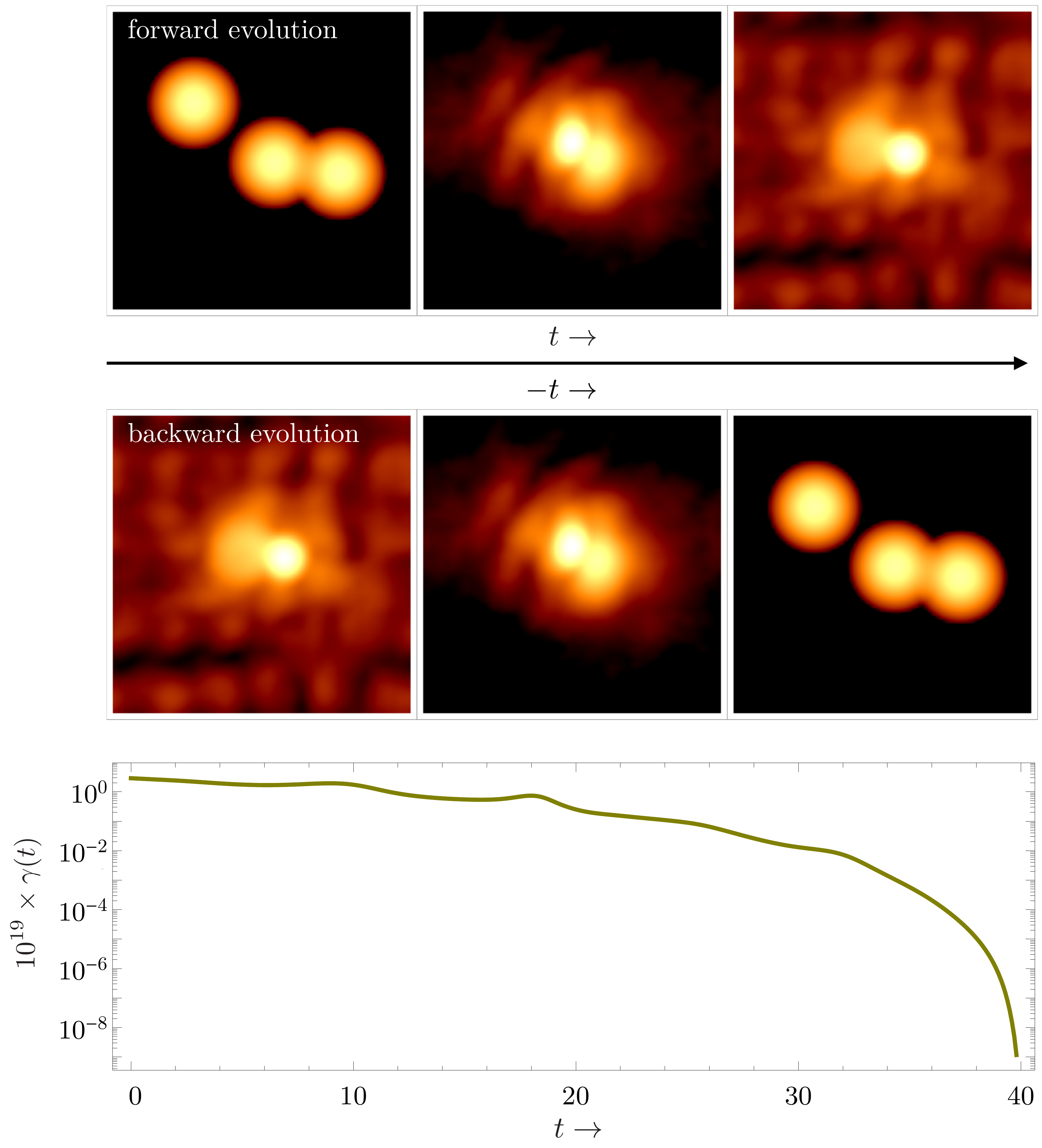}
\caption{Tracking reversibility: Top panel shows snapshots of projected mass density at three instants $t = 0, 13$, and $40$ (upper panel). The self-interaction was chosen to be repulsive ($\lambda=-0.01$). Lower panel include snapshots from the backward evolution at the same instants. The unwinding of the final state to the initial state gives a qualitative proof of reversibility of our algorithm. In the bottom graph we show the asymmetry parameter $\gamma(t)\lesssim 10^{-19}$, which provides a quantitative measure of reversibility.}
      \label{fig:reverse}
\end{figure}

\subsection{Reversibility}\label{sec:reverse}
In order to test reversibility of our algorithm, we define an asymmetry parameter $\gamma$
\begin{align}
    \gamma(t) \equiv \Biggl(\frac{1}{N}\int\mathrm{d}^3x\,\Bigl|\bPsi_{+}(\bx,t) - \bPsi_{-}(\bx,t)\Bigr|^2\Biggr)^{1/2}\,,
\end{align}
where $N$ is the total re-scaled mass (and is already conserved up to machine precision; see the previous subsection), the subscript ``$+$" denotes the forward evolved field starting from some initial condition $\bPsi(\bx,t_i)$ up to a final configuration $\bPsi(\bx,t_f)$, while the subscript ``$-$" denotes the reverse evolved field, starting from the initial condition $\bPsi(\bx,t_f)$ and time-reversed ($\epsilon \rightarrow -\epsilon$ in the simulation). This gives a direct measure of the separation between the forward and backward trajectories.

In~\fref{fig:reverse} we show snapshots from the chosen three soliton collision scenario for repulsive self-interaction along with the asymmetry parameter. We take the distinctive unwinding of the final soliton to the initial 3 soliton state, as well as the smallness of the asymmetry parameter as evidence that our algorithm is indeed reversible.

\subsection{Convergence and accuracy}\label{sec:converge}

Starting from a given initial condition $\bPsi({\bx},0)$ and a time discretization step $\epsilon$, we assume that the simulated field upto time $t$, has the following series representation
\begin{align}\label{eq:series_solution}
    \bPsi_{\epsilon}({\bx},t) = \bPsi_{\rm true}({\bx},t) + \sum^{\infty}_{n=k}{\b c}_n({\bx},t)\epsilon^n\,.
\end{align}
Here $\bPsi_{\rm true}$ is the true solution in the continuum limit $\epsilon \rightarrow 0$, and ${\b c}_n({\bf x},t)$ (independent of $\epsilon$) are the error coefficients at different order in $\epsilon$, with $k$ giving the leading order accumulated error. We then calculate the separation between three trajectories obtained for three different values of $\epsilon$, and construct the following ratio
\begin{align}
\label{eq:C(t)}
    \mathcal{C}(t) = \frac{\Bigl[\int\mathrm{d}^3x|\bPsi_{\epsilon}(t)-\bPsi_{\epsilon/2}(t)|^2\Bigr]^{1/2}}{\Bigl[\int\mathrm{d}^3x|\bPsi_{\epsilon/2}(t)-\bPsi_{\epsilon/3}(t)|^2\Bigr]^{1/2}} = 3^k\left(\frac{2^k - 1}{3^k - 2^k}\right)\, + ...
\end{align}
Here `...' represents $\epsilon$ dependent terms that we neglect. We find $\mathcal{C}(t) \rightarrow 5.4$, implying $k = 2$. Hence the simulated field solution at any given time $t$ is $\mathcal{O}(\epsilon^2)$ away from the true solution. Note that this is one order lower than the truncation error of $\mathcal{O}(\epsilon^3)$ at any time step.\\

We have also checked that the total energy of the system has the same accuracy as that for the scalar case.

\section{Generalizations} \label{sec:extension}
\subsection{Expanding Universe}

With the inclusion of a scale factor $a(t)$ relevant for standard cosmology, the equation of motion reads
\begin{align}\label{eq:masterSP1_expand}
    i\partial_t\bPsi' &= -\frac{1}{2}\frac{\nabla^2}{a^2}\bPsi' + \Phi\,\bPsi' -  \frac{\lambda}{a^3}\Big[(\bPsi'\cdot\bPsi')\,\bPsi^{'\dagger} + 2\,(\bPsi^{'\dagger}\cdot\bPsi')\,\bPsi'\,\Bigr]\,,\nonumber\\
    \frac{\nabla^2}{a^2}\Phi &= \frac{1}{2a^3}\bPsi^{'\dagger}\cdot\bPsi'\,.
\end{align}
Here we have rescaled the field as $\bPsi' = \bPsi\,a^{3/2}$. We only outline the modifications needed in the kick step. With $dt' = a^{-3}\,d{t}$, the corresponding evolution is governed by
\begin{align}\label{eq:vector-kick_ext}
    i\partial_{t'}\psi'_i= [\delta_{ij}(\Phi\,a^3-2\lambda\rho') - \lambda \psi_i^{'*}\psi'_j]\psi'_j\,.
\end{align}
Upon assuming the ansatz $\psi'_j(t') = e^{-i\int^{t'}_{t'_0}\mathrm{d}\tau(\Phi\,a^3 - 2\lambda\rho')}\chi'_j(t')$, we get the same equation as~\eqref{eq:ODE_nd} but with $t$ replaced by $t'$:
\begin{align}
    i\partial_{t'}\chi'_i= -\lambda\chi_i^{'*}\chi'_j\chi'_j\,.
\end{align}
Note that $\rho'$ and $\mathcal{S}'$ are conserved under the above evolution (in terms of $t'$).\footnote{On account of this, also note that $\Phi\propto a$.}

\subsection{Extension to $n$-component Schr\"{o}dinger system \& external potentials}

We now discuss the more general case of a $n$-component Schr\"{o}dinger system. Requiring SO($n$) invariance of the general action
\begin{align}\label{eq:nonrel_action_gen}
    \mathcal{S}_{\rm nr} &= \int \mathrm dt\,\mathrm d^3x\,\Biggl[\frac{i}{2}\bPsi^{\dagger}\cdot\dot{\bPsi} + \mathrm{c.c.} - \frac{1}{2} \nabla\bPsi^{\dagger}\cdot\nabla\bPsi - V_{\rm ext}({\bx})\,\bPsi^{\dagger}\cdot\bPsi - V_{\rm nrel}(\bPsi,\bPsi^\dagger)\Biggr]\,,
\end{align}
the most general form of the quartic self-interaction potential is
\begin{align}
    V_{\rm nrel} = - \frac{\lambda}{2}\Big[2(\bPsi^{\dagger}\cdot\bPsi)^2 + \alpha(\bPsi\cdot\bPsi)\,(\bPsi^{\dagger}\cdot\bPsi^{\dagger}) \Bigr]\,,
\end{align}
where $\lambda \in \mathbb{R}$ and $\alpha \in \mathbb{R}$.\footnote{The scalar case $n=1$ is obtained by absorbing $\alpha+2$ into $\lambda$, while for the vector case $\alpha = 1$.} In the above action, $V_{\rm ext}({\bf x})$ is any general potential (including the gravitational potential).

Apart from total mass, linear and angular momentum, and energy, the total `isospin' in this case is the charge associated with the internal SO($n$) symmetry. With $L_{a}$ as the skew-symmetric $n \times n$ matrices/generators (where the integer index $a \in [1,n(n-1)/2]$), it takes the general form\footnote{For $n = 3$, $L_i$ can be characterized by Levi-Civita symbols, $[L_{i}]_{jk} = \epsilon_{ijk}$, giving the usual spin density $\mathcal{S}_i$.}
\begin{align}
    I_a = \int\mathrm{d}^3x\,\mathcal{I}_a \qquad {\rm where} \qquad \mathcal{I}_a = i\psi_i\,[L_a]_{ij}\,\psi^{\ast}_j = {\rm isospin\,density}\,.
\end{align}
Similar to the case of vectors, one way to handle the kick step is to add and subtract $\alpha\,\rho^2$ in the above potential to give $V_{\rm nrel} = - (\lambda/2)\Big[(2+\alpha)\rho^2 - \alpha\,\mathcal{I}^2 \Bigr]$, which ultimately renders the following kick evolution for $\psi$
\begin{align}
    \psi_i(t) = e^{-i(\Phi - (2+\alpha)\lambda\rho)(t-t_0)}\,[e^{-i\alpha\lambda\,\mathcal{I}\cdot L (t-t_0)}]_{ij}\psi_j(t_0)\,.
\end{align}
In order to avoid the non-trivial exponentiation of the matrix $\mathcal{I}\cdot L$, we can instead use our trick of decoupling the different Cartesian components. With this approach, the algorithm for evolving the $n$-component system is exactly the same as developed in this paper. That is, the same steps as outlined in section~\ref{sec:vector_algo_summary}, with $\Phi \rightarrow V_{\rm ext}$ and the factor of $\lambda$ replaced by $\alpha\lambda$ in $\mathcal{U}$ (c.f.~\eqref{eq:kick_operator}). The only difference is that now we have $n$ fields to evolve instead of $3$.

\section{Summary}
\label{Sec:sum_disc}

We have devised an algorithm for evolving multicomponent Schr\"{o}dinger systems obeying SO($n$) symmetry and containing both gravitational and non-gravitational self-interactions, in particular (iso-)spin-spin interactions that mix the different field components. The crucial aspect of our algorithm is the kick step, arising due to all the interactions. Approximate extensions of existing scalar algorithms with self-interactions or multicomponent algorithms without (iso-)spin-spin interactions to our system can lead to a loss of spin, mass conservation and also reversibility. An analytic solution for the kick step containing (iso-)spin-spin interactions mitigates these issues.

We provided such an analytic solution for the kick step. We first showed that different field components can be decoupled in their evolution. This was made possible by noticing that the magnitude of $\psi_i\psi_i$ is a constant of motion throughout the kick step, since both the number density and (iso-)spin density remain constant. Owing to this decoupling, we get a reduced $2$-level system in which only the real and imaginary parts of each of the field components remain coupled, with the evolution of each component being the exactly the same. We are then able to find an exact analytical solution for this reduced coupled system. Equipped with this analytical solution, we developed a split-step Fourier algorithm that involves drift-kick-drift set of operations, in analogy to the scalar case. Our algorithm is time reversible, unitary and spin/iso-spin conserving (up to machine precision). 

For concreteness, we introduced the above algorithm first for a $3$-component system (vector/ spin-$1$ field).\footnote{Compared to the earlier version of this paper, in this version we also provide a separate algorithm for the $3$-component case where the different field components are coupled in their evolution. This is similar to the algorithm used in the BEC literature~\cite{PhysRevE.93.053309}, but in the spin basis (as opposed to our Cartesian basis). While this algorithm is straightforward for small $n$, the extension to the $n$-component case which we formally provide in section~\ref{sec:extension}, can become cumbersome when $n$ is large.} To demonstrate the fidelity of our algorithm, we showed that for sample simulations (with gravity + attractive or repulsive self-interactions) that total mass and total spin are conserved up to machine precision, and the algorithm maintains time reversibility. The overall accuracy of the algorithm, as compared to the true (continuous) evolution, is $\mathcal{O}(\epsilon^2)$ and is therefore on par with the usually employed split-step Fourier algorithm for a single scalar field, or multiple scalar fields without the aforementioned spin-spin self-interaction.

Our algorithm is general and flexible enough for a wide range of applications in astrophysics, cosmology and condensed-matter physics. We have provided generalizations of the algorithm to include an expanding cosmological background, and external potentials in laboratory systems, and an arbitrary number of components (respecting SO($n$) symmetry). The computational cost of including self-interactions is only order unity larger than the case without them.

With this paper, we make the code (written in Python) for a multicomponent Schr{\"o}dinger system with/without gravity + with/without self-interactions, publicly available at~\href{https://github.com/mudit-jain90/i-SPin}{GitHub}.

\acknowledgments

We would like to especially thank Wisha Wanichwecharungruang (Rice U.) for help in testing various aspects of an earlier version of the algorithm presented in this paper. We would also like to thank Dorian Amaral for helpful discussions surrounding the NumPy library in Python. We acknowledge Rohith Karur's involvement and help in the initial stages of the project, and thank Andrew Long, Philip Mocz, Jonathan Thomas and Han Pu for discussions. MA and MJ are partly supported by a DOE grant DE-SC0021619.

\bibliographystyle{utphys}
\bibliography{reference}
\appendix

\newpage
\section{Appendix}\label{sec:sim}
\subsection{Non-relativistic limit}
\label{sec:app_nrlimit}
A (dark) massive spin-$1$ field $W_\mu$ minimally coupled to gravity and with non-gravitational self-interactions, is described by the following effective action:
\begin{align}\label{eq:rel_action}
    S &= \int\mathrm{d}^4x\,\sqrt{-g}\Bigl[-\frac{1}{4c}\mathcal{G}^{\mu\nu}\mathcal{G}_{\mu\nu} + \frac{1}{2}\frac{m^2c^2}{\hbar^2}\,W_{\mu}W^{\mu} + V_{\rm rel}(W_\mu W^\mu) - \frac{c^3}{16\pi G}R + ...\Bigr].
\end{align}
Here $\mathcal{G}_{\mu\nu}=\partial_\mu W_\nu-\partial_\nu W_\mu$, and the `$...$' represents the Standard Model Lagrangian and other possible dark sector(s). The parameter $m$ is the mass of the vector boson, and $V_{\rm rel}$ contains self-interactions of the vector field, arising on account of some ultraviolet physics. We shall be interested in the leading order (dimension 4) operator describing a quartic interaction, which takes the form
\begin{equation}
    V_{\rm rel}=\frac{\lambda}{4}(W_\mu W^\mu)^2 + \hdots\,,
\end{equation}
Without loss of generality, the spatial part of the (real-valued) vector field  $\bW$ can be represented in terms of a complex vector $\bPsi$ as
\begin{align}
\label{nonrelativistic_expansion}
&\bW(t,\bx) \equiv \hbar\sqrt{\frac{2}{mc}}\Re\left[\bPsi(t,\b x) e^{-imc^2t/\hbar}\right],
\end{align}
where $\bPsi$ has dimensions of $[\textrm{length}]^{-3/2}$. Similarly, $W_0(t,\bx)\equiv\hbar\sqrt{2/mc}\,\Re\left[ \psi_0(t,\b x) e^{-imc^2t/\hbar}\right]$. We are interested in the non-relativistic behavior of the vector field where the spatial variation in the field is slow compared to the Compton scale $\lambda_m=\hbar/mc$. For capturing this non-relativistic behaviour, we insert the above form of $W_\mu$ and drop all the second time derivative terms acting on $\bPsi$ as well as terms containing the fast oscillating pieces of the form $e^{\pm i n mc^2t/\hbar}$. We also restrict ourselves to Newtonian gravity.\footnote{To leading order in the non-relativistic limit, $|\nabla| \sim k \ll mc/\hbar$, $\psi_0 = i\hbar\,\nabla\cdot\bPsi/mc$. That is, $\psi_0$ is determined from $\bPsi$ and subdominant compared to $\bPsi$.} With these considerations, we arrive at the nonrelativistic action for the slowly varying part of the vector field and Newtonian gravity:
\begin{align}
    \mathcal{S}_{\rm nr} &= \int \mathrm dt\,\mathrm d^3x\,\Biggl[\frac{i\hbar}{2}\bPsi^{\dagger}\dot{\bPsi} + \mathrm{c.c.} - \frac{\hbar^2}{2m} \nabla\bPsi^{\dagger}\cdot\nabla\bPsi + \frac{1}{8\pi G}\,\Phi\nabla^2\Phi - m\,\Phi\,\bPsi^{\dagger}\cdot\bPsi - V_{\rm nrel}(\bPsi,\bPsi^\dagger)\Biggr]\nonumber\\
    &\qquad {\rm with } \qquad V_{\rm nrel}(\bPsi^\dagger,\bPsi) = - \frac{\lambda(\hbar c)^3}{8(mc^2)^2}\Big[(\bPsi\cdot\bPsi)\,(\bPsi^{\dagger}\cdot\bPsi^{\dagger}) + 2\,(\bPsi^{\dagger}\cdot\bPsi)^2\,\Bigr]\,.
\end{align}
For the case of only gravitational interactions, the nonrelativistic limit was derived earlier in \cite{Adshead:2021kvl}. For massive spin-2 case, see~\cite{Aoki:2017ixz}. For a generalization to the spin-$s$ case including spin-$1$ and spin-$2$ cases, see \cite{Jain:2021pnk}. Going further, relativistic corrections to this multicomponent system should be investigated. See~\cite{Salehian:2021khb} for single-component/scalar case. 

\subsection{Fluid and Conservation Equations}
\label{sec:app_fluid_conserve_eqns}

We can also transform our multicomponent SP system eq.~\eqref{eq:masterSP1} into a set of three, coupled fluid equations (following the Madelung transform commonly used in SDM \cite{1927ZPhy...40..322M}). With the following field re-definition, $\psi_j=\sqrt{\varrho_j/m}\,e^{i\Theta_j}$, and the velocity $\bm{u}_i \equiv \hbar\nabla \Theta_i/m$, we have (for $j=1,2,3$):
\begin{align}
    &\frac{\partial\varrho_j}{\partial t} + \nabla\cdot(\varrho_j \bm{u}_j) = R^{\rm SI}_j\rho_j\,,\\
    &\frac{\partial\bm{u}_j}{\partial t} + (\bm{u}_j\cdot{\nabla})\bm{u}_j = \frac{1}{m}\nabla(Q_j +Q^{\rm SI}_j- m\Phi),
\end{align}
where $\varrho_j=m\rho_j$ is the mass density of each component and
\Beq
R^{\rm SI}_j&=\frac{\lambda\hbar^2}{2m^3c}\sum_{i=1}^3\varrho_i\sin\left[2(\Theta_j-\Theta_i)\right]\,,\\
Q_j &= \frac{\hbar^2}{2m}\frac{\nabla^2\sqrt{\varrho_j}}{\sqrt{\varrho_j}}\,,\qquad Q_j^{\rm SI}=\frac{\lambda \hbar^3}{2m^3 c}\sum_{i=1}^3\varrho_i\left(1+\frac{1}{2}\cos[2(\Theta_j-\Theta_i)]\right).
\Eeq
Note that while the mass (and particle number) of individual components is not conserved in presence of self-interactions, the total mass (and number) is conserved. That is, $\sum_{j=1}^3R_j^{\rm SI}\varrho_j=0.$ Furthermore, the potential $Q_j^{\rm SI}$ is present only when we include non-gravitational self-interactions, whereas $Q_j$ is present with gravity alone.

For future reference, we provide the continuity equations for the number and spin densities also. Defining the general Schr\"{o}dinger current tensor
 \begin{align}\label{eq:tensor_current}
     \bm{\mathcal{J}}_{jk} \equiv i\frac{\hbar}{2m}\left[\psi_{j}\nabla\psi^{\ast}_{k} - \psi^{\ast}_{k}\nabla\psi_{j}\right]\,,
 \end{align}
 the continuity equations are
 \begin{align}\label{eq:cont_eqns}
 \partial_t \mathcal{N} +\nabla\cdot \bm{\mathcal{J}}_{ll}=0\,,\qquad {\rm and} \qquad
 \partial_t \mathcal{S}_i + i\hbar\,\epsilon_{ijk}\,\nabla\cdot\bm{\mathcal{J}}_{jk}=0\,.
 \end{align}

\subsection{Lowest energy soliton solutions}\label{sec:solitons}

In order to obtain soliton states, we can extremize the energy functional at a fixed particle number $N_0$, i.e. extremize the quantity $H+\mu c^2(N-N_0)$ where $\mu c^2$ can be thought of as the chemical potential. This is equivalent to assuming the ansatz $\bPsi(\bx,t) = e^{i\mu c^2t/\hbar}\,\bepsilon^{(s)}_{\hat{n}}\,f_s(\bx)$, where in order to get the lowest energy solutions, $f_s(\bx)$ is a radially symmetric function and $\bepsilon^{(s)}_{\hat{n}}$ is a spatially independent (and normalized) polarization vector characterizing the polarization of the soliton along the $\hat{n}$ direction. The label $s$ corresponds to the spin multiplicity, which takes the values $0$ or $1$ for linearly polarized and circularly polarized soliton respectively. For example, $\bepsilon^{(0)}_{\hat{z}} = (0,0,1)$ and $\bepsilon^{(\pm 1)}_{\hat{z}} = (1,\pm i,0)/\sqrt{2}$. See~\cite{Jain:2021pnk,Jain:2022kwq} for details. This renders the following time-independent Schr\"{o}dinger-Poisson system
\begin{align}\label{eq:TISP}
    -\mu c^2f_s(\bx) &= -\frac{\hbar^2}{2m}\nabla^2f_s(\bx) + m\,\Phi(\bx)\,f_s(\bx) - \frac{\lambda\hbar^3}{4m^2c}\left(3-s\right)\,f^3_s(\bx)\nonumber\\
    \nabla^2\Phi(\bx) &= 4\pi G m\,f_s^2(\bx)\,.
\end{align}
The solutions to the above set of equations, for different $\mu$ values (and given $m$ and $\lambda$), can be obtained by numerical shooting method. Note that there is a scaling symmetry in the system where different soliton solutions (with different values of $\mu$) for a given $\lambda$ and $m$, can be mapped to solitons with different $\lambda$ and same $\mu$ and $m$. That is, once a soliton solution is obtained for a $\mu$, another soliton solution for a different $\mu'$, can be obtained by re-scaling the fields, space, and $\lambda$ as $\Phi \rightarrow (\mu'/\mu)\Phi$, $f_s \rightarrow (\mu'/\mu)f_s$, $\bx \rightarrow (\mu/\mu')^{1/2} \bx$, and $\lambda \rightarrow (\mu'/\mu)\lambda$.

\begin{figure}[t!]
    \centering
      \includegraphics[width=6.5in]{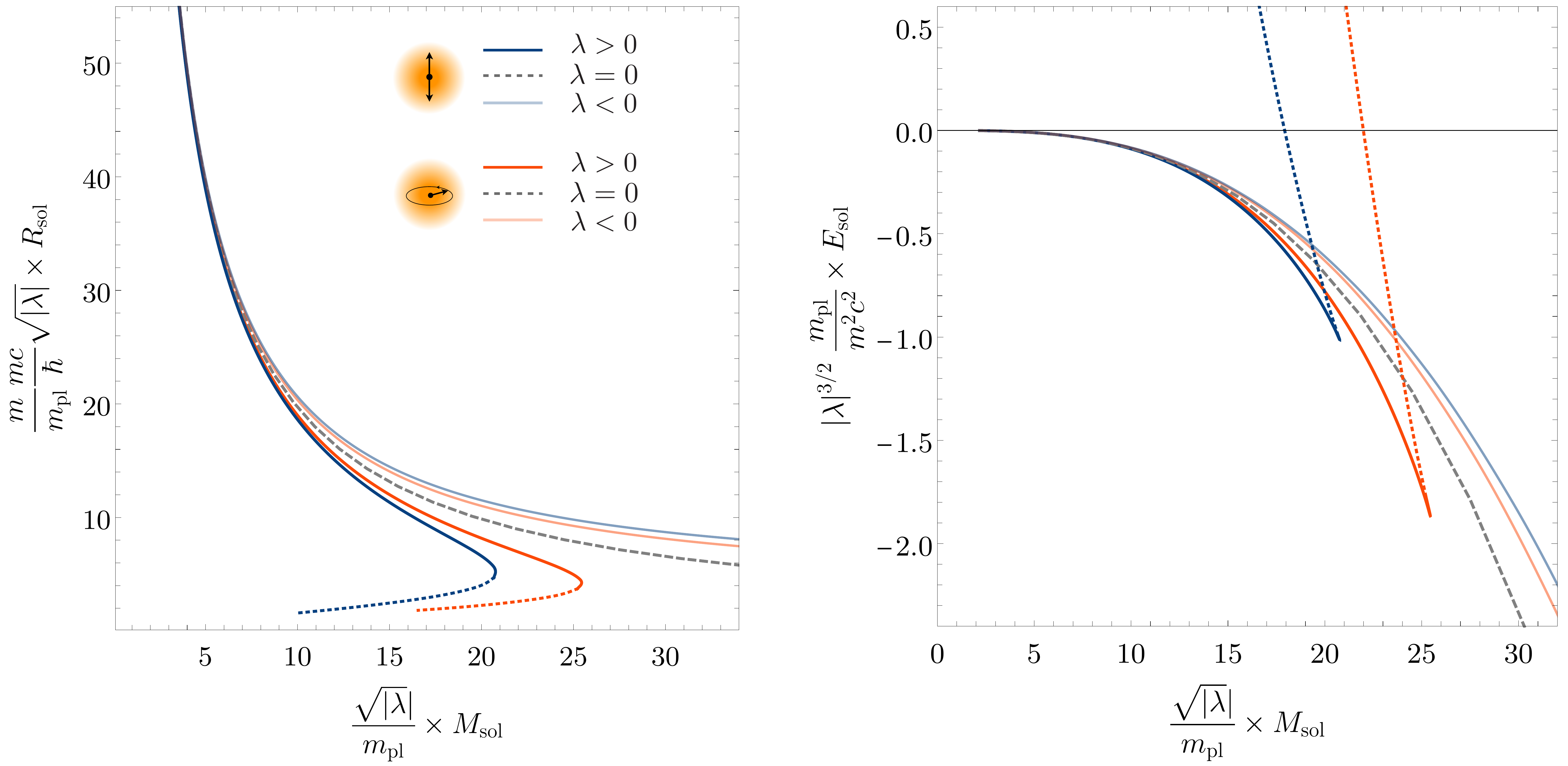}
\caption{\small{Radius vs mass (left panel), and energy vs mass (right panel) for vector solitons with self-interactions.} The dotted black curve is the limit as $\lambda \rightarrow 0$ (without any factors of $\lambda$ appearing in the re-scaling of axes). The curves lying above and below the dotted black curve on the left panel, are for repulsive ($\lambda < 0$) and attractive ($\lambda > 0$) self-interactions respectively. While on the right panel the curves lying above and below the dotted black curve, are for attractive and repulsive self-interactions respectively. }
      \label{fig:soliton.curves}
\end{figure}

\fref{fig:soliton.curves} shows the two possible families of solitons, corresponding to linear ($s=0$), and circular polarization ($s=1$), for both attractive ($\lambda > 0$)  and repulsive ($\lambda < 0$) interactions.\footnote{See~\cite{Jain:2022kwq} for a discussion of these different families of solitons arising in Abelian and non-Abelian Higgs model.} Here $M_{\rm sol}$ and $E_{\rm sol}$ are the mass and energy (excluding rest mass) of the soliton, and  $R_{\rm sol}$ is the radius containing $95\%$ of the soliton mass. Solitons where attractive self-interaction balances the gradient pressure (with gravity being sub-dominant) are unstable~\cite{Schiappacasse:2017ham,Jain:2022kwq} and lie on the colored dashed curves.

To highlight the difference between the non self-interaction case ($\lambda = 0$) for which $R_{\rm sol} \propto 1/M_{\rm sol}$ and $E_{\rm sol} \propto M_{\rm sol}^3$, we have also plotted the respective curves in dashed gray. From the right panel in Fig.~\ref{fig:soliton.curves}, note that solitons with the same mass have different energies for different polarizations (spin) when non-gravitational self-interactions are present.\footnote{In the case of no self-interactions and leading non-relativistic limit, there are infinitely many degenerate solitons, with total spin ranging anywhere from $0$ to $\hbar N$, called fractionally polarized solitons~\cite{Jain:2021pnk}.} For a fixed $|\lambda|$ and mass $M_{\rm sol}$, linearly polarized soliton is the lower energy state when the interaction is attractive ($\lambda>0$). Whereas for repulsive self-interaction ($\lambda < 0$), it is the circularly polarized soliton that is lower in energy.

\subsection{Explicit expression for the evolution operator $\mathcal{U}$}
\label{sec:app:Uexplicit}

With $\mathcal{S} \equiv (\bm{\mathcal{S}}\cdot\bm{\mathcal{S}})^{1/2}$ where $\bm{\mathcal{S}}$ is the spin density, $r = |\bPsi\cdot\bPsi|$, and ${\rm sinc}(x) = \sin(x)/x$ as the sinc function, the 4 different components of $\mathcal{U}$ are
\begin{align}
    \mathcal{U}(t-t_0)_{11} &= \cos(\lambda\rho(t-t_0))\cos(\lambda \mathcal{S}(t-t_0))\nonumber\\
    &\;\;\; + \lambda(t-t_0)\,\sink(\lambda \mathcal{S}(t-t_0))\Bigl[\rho \sin(\lambda\rho(t-t_0)) - r \sin(\theta + \lambda\rho(t-t_0))\Bigr]\nonumber\\
    \mathcal{U}(t-t_0)_{12} &= -\sin(\lambda\rho(t-t_0))\cos(\lambda \mathcal{S}(t-t_0))\nonumber\\
    &\;\;\; + \lambda(t-t_0)\,\sink(\lambda \mathcal{S}(t-t_0))\Bigl[\rho \cos(\lambda\rho(t-t_0)) + r \cos(\theta + \lambda\rho(t-t_0))\Bigr]\nonumber\\
    \mathcal{U}(t-t_0)_{21} &= \sin(\lambda\rho(t-t_0))\cos(\lambda \mathcal{S}(t-t_0))\nonumber\\
    &\;\;\; + \lambda(t-t_0)\,\sink(\lambda \mathcal{S}(t-t_0))\Bigl[-\rho \cos(\lambda\rho(t-t_0)) + r \cos(\theta + \lambda\rho(t-t_0))\Bigr]\nonumber\\
    \mathcal{U}(t-t_0)_{22} &= \cos(\lambda\rho(t-t_0))\cos(\lambda \mathcal{S}(t-t_0))\nonumber\\
    &\;\;\; + \lambda(t-t_0)\,\sink(\lambda \mathcal{S}(t-t_0))\Bigl[\rho \sin(\lambda\rho(t-t_0)) + r \sin(\theta + \lambda\rho(t-t_0))\Bigr]\,.
\end{align}

\end{document}